\documentclass[fleqn,usenatbib]{mnras}

\usepackage{newtxtext,newtxmath}

\usepackage[T1]{fontenc}
\usepackage{ae,aecompl}

\usepackage{graphicx}	
\usepackage{amsmath}	
\usepackage{amssymb}	
\usepackage{ulem}

\newcommand\asec{$''$}

\title[Quiet-sun turbulence]{Analysis of quiet-sun turbulence on the basis of SDO/HMI and Goode Solar Telescope data}

\author[V.I. Abramenko and V.B. Yurchyshyn]
{Valentina I. Abramenko$^{1}$ \thanks{E-mail: vabramenko@gmail.com (VIA)}
and Vasyl B. Yurchyshyn$^{2}$ \\
	$^{1}$Crimean Astrophysical Observatory, p/o Nauchny, Crimea, 298409, Russia\\
	$^{2}$Big Bear Solar Observatory, New Jersey Institute of Technology, Big Bear City, CA, 92314, USA\\
}

\date{Accepted XXX. Received YYY; in original form ZZZ}

\pubyear{2020}

\begin{document}
\label{firstpage}
\pagerange{\pageref{firstpage}--\pageref{lastpage}}
\maketitle

\begin{abstract}

We analysed line-of-sight magnetic fields and magnetic power spectra of an undisturbed photosphere using magnetograms acquired by the Helioseismic and Magnetic Imager (HMI) on-board the Solar Dynamic Observatory (SDO) and the Near InfraRed Imaging Spectrapolarimeter (NIRIS) operating at the Goode Solar Telescope (GST) of the Big Bear Solar Observatory. In the NIRIS data revealed the presence of thin flux tubes of 200-400~km in diameter and of field strength of 1000-2000~G. The HMI power spectra determined for a coronal hole, a quiet sun and a plage areas exhibit the same spectral index of -1 on a broad range of spatial scales from 10-20~Mm down to 2.4~Mm. This implies that the same mechanism(s) of magnetic field generation operate everywhere in the undisturbed photosphere. The most plausible one is the local turbulent dynamo. When compared to the HMI spectra, the -1.2 slope of the NIRIS spectrum appears to be more extended into the short spatial range until the cutoff at 0.8-0.9~Mm, after which it continues with a steeper slope of -2.2. Comparison of the observed and Kolmogorov-type spectra allowed us to infer that the Kolmogorov turbulent cascade cannot account for more than 35\% of the total magnetic energy observed in the scale range of 3.5-0.3~Mm. The energy excess can be attributed to other mechanisms of field generation such as the local turbulent dynamo and magnetic super-diffusivity observed in an undisturbed photosphere that can slow down the rate of the Kolmogorov cascade leading to a shallower resulting spectrum.

\end{abstract}




\section{Introduction}

The undisturbed photosphere outside active regions (ARs) occupies up to 80\% of the solar surface and hosts a significant amount of magnetic flux in form of small-scale magnetic elements continuously buffeted and displaced by turbulent plasma flows, which ultimately leads to flux cancellation and submergence. Physical processes in the hierarchy of magnetic elements play a key role in elaboration of an adequate scenario of solar dynamo.

The magnetic power spectrum analysis is a powerful tool for exploring photospheric turbulence in quiet Sun (QS) and it was extensively used since early 70s.  \citet[][]{Nakagawa1973} suggested to consider the transport of small-scale magnetic elements in the photosphere as a passive response of the vertical component of the magnetic field to turbulent plasma motions. This allowed authors to apply various analysis techniques of theory of turbulence to solar magnetograms of active and quiet regions. Using 2\asec.5 resolution data from Kitt Peak National Observatory, they reported a $E(k) \sim k^{-1}$ spectrum for a typical AR and a $k^{-0.3}$ spectrum for a QS area on spatial scales larger than 10~Mm at which both spectra show a cutoff with much steeper slopes afterwards. More magnetic power spectra studies followed \citep[e.g.,][]{Knobloch1981,Lee1997}, also see \cite{Abramenko2001} for a review of early publications. It was pointed out in \citet{Abramenko2001} that as the spatial resolution improved the position of the spectrum cutoff has shifted along the spectrum toward smaller scales.  \citet{Abramenko2001} also showed that the cutoff is observed on a scale, $r_0$, corresponding to the approximately triple resolution limit of the instrument, $r_{res}$, while data correction using the modulation transfer function (MTF) shifts the cutoff to $\sim(1.5-2)r_{res}$. Using magnetic field data from the Big Bear Solar Observatory (BBSO) videomagnetograph with a pixel scale of 0\asec.76$\times$0\asec.60 \citet{Abramenko2001} obtained a QS spectrum $k^{-1.3}$ within a linear range of (9-3)~Mm without the MTF correction. The lower limit of 3~Mm exceeds the spatial resolution of the data ($r_{res}$ = 2\asec) by 2.2 times.  

Launch of \textit{Hinode} enabled further progress in studying small scale magnetic fields of QS areas.    
In a set of fundamental publications by \citet{Stenflo2010,Stenflo2011,Stenflo2012} on the nature of the quiet sun magnetic field,  the magnetic power spectrum is in the centre of attention. A Hinode SOT/SP ``deep mode'' data set recorded in a QS area near the disk centre on February 27, 2007 \citep{Lites2008} was used by \citet{Stenflo2010} to calculate a 1D power spectrum of the vertical flux density on spatial scales ranging from 100~Mm down to the limit of the Hinode resolution of 0.232~Mm. After MTF-correction the spectrum showed excess of energy on scales below 2~Mm as compared to the original spectrum. An analytical spectrum extending beyond the Hinode resolution was derived from an analytical study of the collapsed flux tubes statistics considering the hidden flux based on the Hanle effect \citep{Stenflo2010, Stenflo2011}. The author argued that the intrinsic 1D power spectrum contains two inertial ranges, both with the slope of -5/3. The first range covers the observable scales, while the second one spans an interval from approximately 100~km down to the Ohmic dissipation scale of about 25~m. One of the key points in this study is the existence of an ensemble of collapsed flux tubes of size of 10-70~km, which may contribute to the spectrum mostly on scales of 10-150~km thus producing a local power enhancement  that breaks  the linear -5/3 slope of the spectrum. The magnetic flux on smaller scales is thought to be represented by a hidden flux of tangled fields. The author does not support the idea of parallel existence of local and global dynamos, instead, it is suggested that it is ``the self-sustaining processes that maintain the magneto-convective spectrum in the stratified Sun'' \citep{Stenflo2012}. 

\citet{Katsukawa2012} analysed a set of Hinode SOT/SP QS magnetograms obtained between November 2006 and December 2007 paying special attention to correcting the spectra using different forms of the MTF. One of them was based on an assumption that the wavefront error derived for the Hinode/BFI instrument can be applied to the SP instrument. As a result, they found that the energy of the corrected spectrum exceeds that of the original one on scales below $\sim$10~Mm. The data shown in their Figure 2 suggest that the slope of the original spectrum determined on the range of 2-15~Mm is close to $-1/2$ or slightly steeper. The MTF-corrected spectrum in this range was found to be nearly flat and it only became steeper ($k^{-1.4}$) on sub-granular scales ($<$1~Mm). These authors interpreted the $k^{-1.4}$ spectrum as a manifestation of magnetic energy transferred to these scales from both large and small scales. 

\citet{Danilovic2016} applied a sophisticated technique to compare QS power spectra derived from Honode SOT/SP observations and from MHD simulations of local dynamo. This Hinode data set was previously analysed and described in \citep{Lites2008}, while one simulated set was generated by the MURaM code \citep{Vogler2005} and the other one was local dynamo simulations by \citet{Rempel2014}. A 2D inversion technique \citep{vanNoort2012} was used to degrade the simulated data to Hinode SOT/SP resolution, while the observed data were MTF corrected. It was shown that MTF correction increases the observed power over the entire analyzed spectral range from 60~Mm down to the Hinode diffraction limit of 0.232~Mm and the effect of correction becomes stronger on smaller scales. These authors also reported that the observed kinetic and magnetic power spectra have slopes similar to those derived from simulations. These findings may be interpreted as evidence of local dynamo operating in the QS photosphere. 

It is worth noting that the three above studies used data acquired by the same instrument during a relatively short time interval (2006-2007) and were processed by the same routine, $sp{\_}prep$ (except the set used in \citet{Stenflo2012}), however they all followed different approach to the MTF correction, which affected the outcome. The reason is that the instrumental image degradation leads to unavoidable loss of information. All attempts to recover this information strongly relay on a set of assumptions adopted to construct an MTF \cite[e.g.,][]{Suematsu2018}. The situation is even more severe for ground-based instruments mainly because of the MTF is continuously changing due to seeing conditions. \citet{Abramenko2001} showed that on scales larger than approximately the triple resolution limit of a telescope (the coefficient may range from 2.7 to 3.3 with the average of 3.0 depending on the instrument), the influence of the MTF can be negligible. This consideration motivated us to avoid the MTF-correction and to restrict our analysis to a spatial range $>3r_{res}$. Earlier \citet{Abramenko2012} suggested that as the resolution of solar instruments improves, the $k^{-1}$ magnetic power spectrum should extend toward the smaller scales, with the slope remaining the same. High resolution measurements of the magnetic field obtained with BBSO's Goode Solar Telescope (GST) motivated us to further explore this tendency. On the other hand, to the best of our knowledge, no comprehensive study of the magnetic power spectrum in an undisturbed photosphere based on data from the Helioseismic and Magnetic Imager (HMI) on-board Solar Dynamic Observatory (SDO) were undertaken as yet. The comparison of power spectra derived from these two instruments is the one of the goals of this study. 

\section{Data and Method}

\begin{figure*}
 	\includegraphics[width=\linewidth]{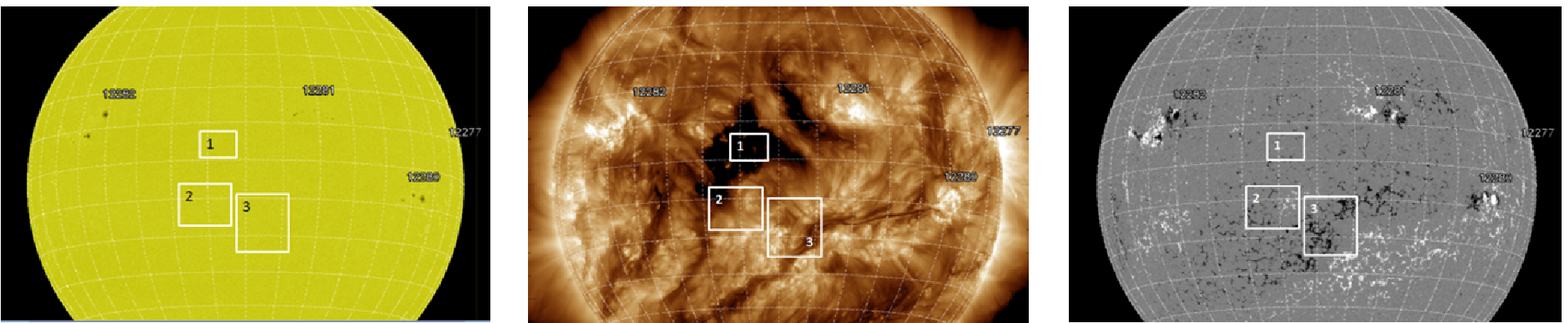}
 	\caption{\sf Three types of undisturbed photosphere analysed in this study. Boxes 1, 2, and 3 enclose a fraction of a coronal hole, a quiet Sun area, and a plage region, correspondingly. The data were obtained on 10 February, 2015. Left, middle, and right panels show an HMI 6173~\AA\ image, an AIA 193~\AA\ image, and an HMI line-of-sight magnetogram scaled between $\pm$200~G, correspondingly. }
 	\label{fig-1}  
 \end{figure*}

We analysed three sets of photospheric magnetograms listed in  Table 1. The first set is comprised of 11 consecutive full disc HMI line-of-sight magnetograms (hmi.M-720s series) taken in Fe I 6173.3~\AA\ spectral line on 2015 February 10 with the spatial resolution of 1\asec (pixel size of 0\asec.5), cadence of 12~min \citep{Schou2012} and noise level of about 6~Mx cm$^{-2}$ \citep{Liu2012}. Three QS regions near the disk centre (enclosed by boxes in Figure \ref{fig-1}) representing different type of undisturbed photosphere were selected for the study. Box 1 encloses a fraction of a coronal hole (CH area), box 2 represents a QS area of closed magnetic field lines, and box 3 is a plage region formed on the remnants of NOAA active region 12259. For each selected area we derived the average magnetic power spectrum by averaging 11 individual spectra separately calculated for each magnetogram in the data set (Figure \ref{fig-2}).

 The second data set contains HMI/hmi.M-720s series data obtained on 2017 June 19 from 18:00~UT fo 20:00~UT (11 magnetogrms). The utilized area of 930$\times$530 HMI pixels covered a vast QS region of the very weak magnetic field at the solar disc center. The corresponding averaged over 11 magnetograms spectrum is shown in Figure \ref{fig-2}. The choice of the date, time interval and the analyzed HMI area was defined by the cotemporary data set acquired by the BBSO's Near InfraRed Imaging Spectrapolarimeter \citep[NIRIS,][]{Cao2012}. 	  

This third NIRIS data set was acquired on 2017 June 19 between 18:37:46~UT and 19:22:50~UT (Figure \ref{fig-3}, left panel). The same data set was used in a study of chromospheric spicules by \cite{Samanta2019}. The measurements were made using adaptive optics corrected light, a dual Fabry-Perot etalon, and a 2K$\times$2K HgCdTe Helium cooled Teledyne camera. Two polarization states are simultaneously imaged side-by-side on a 1024$\times$1024 pixel area each, using a dual beam system that provides an 85\asec round FOV with image scale of 0\asec.083 per pixel. The measurements were performed using the Fe I 15650~\AA\ doublet with a bandpass of 0.1~\AA\ and a rotating 0.35 wave plate that allowed us to sample 16 phase angles at each of more than 60 line positions at a cadence of 30 s per one full spectroscopic measurement (full-Stokes I, Q, U, V). The Fe I 15650~\AA\ Stokes data were corrected for polarization effect and inverted using a Milne-Eddington (ME) inversion approach adopted for NIRIS data. 

The magnetic power spectrum calculations were performed using the approach described in  \citet{Abramenko2001} and further refined in \citet{Abramenko2005}. Briefly, the routine can be summarized as follows. The power spectrum of an n-dimensional structure $u(\mathbf{r})$ is defined via its Fourier transform  $U(\mathbf{k})$ as:
\begin{equation}
F(\mathbf{k}) =  \frac{1}{2}\vert U(\mathbf{k}) \vert ^2,
\label{eq1}
\end{equation}  
where $\mathbf{k}$ is an n-dimensional wave vector. Frequently, a less informative statistical description is used such as an 1D spectrum mainly because it is the 1D spectrum, $E(k)$, that is studied in theory of local isotropic turbulence. Many theoretical inferences were derived from 1D spectrum analysis, including the Kolmogorov-type scaling law of -5/3 \citep{Monin1975}. In order to obtain a 1D spectrum, function $F(\mathbf{k})$ needs to be integrated over the area encircled by two concentric spheres (circles for n=2) separated by a small interval $\Delta \mathbf{k}$:
\begin{equation}
E(k) =  \int_{|\mathbf{k}|=k} F(\mathbf{k}) dS(\mathbf{k}),
\label{eq2}
\end{equation}
where $dS(\mathbf{k})$ is an element of the area of the sphere $|\mathbf{k}|=k$. This technique was applied in \citet{AbramenkoYur2010, McAteer2010, Katsukawa2012, Kutsenko2019}.

As we have shown in  \citet{Abramenko2001}, the most reliable linear interval to determine the power index, $\alpha$, of the spectrum, $E(k) \sim k^{\alpha}$ calculated for solar magnetograms is an interval from 10 to 3~Mm. On large scales, large sunspots, if present in  the FOV under study, might drastically affect the shape of a spectrum by injecting significant energy on large spatial scales. On smaller scales below $\sim$3~Mm, the MTF effect might lower the measured energy density. This strategy was \textbf{first }elaborated for analyzing active regions using MDI data and was used in \citet{AbramenkoYur2010} and \citet{McAteer2010}. A modification of this approach is needed when analysing photospheric magnetograms that do not include sunspots.

A real turbulent spectrum consists of three main ranges \citep{Monin1975}. On large scales there is a nearly flat (or slightly concave) range of energy injection. It is followed by the so called inertial power-law range, where the energy injected at large scales is transferred without loss to smaller spatial scales. This range is the most useful for diagnostics of turbulence. On small scales, where the energy dissipation becomes significant, the third, dissipation range is observed with a much steeper slope. The real inertial range might be broader than what our data allow us to estimate. Our aim is to determine a portion of the spectrum inside the inertial range to calculate the most reliable slope.    

In general, the transition from the energy injection range to the inertial range is very smooth and the magnetic power spectra derived from both HMI data sets (Figure 
\ref{fig-2}) confirm this. 
 A location of the transition depends on the size of  large-scale features of the field under study \citep{Monin1975}. Thus, for a structure with rather small dominating features, the energy injection range is rather broad and smooth and the transition to the inertial range occurs on rather small scales. In the case when strong large features are present (for example, the super-granulation cells, see Figure \ref{fig-1}, boxes 2 and 3), the inertial range extents toward larger scales and local maxima can appear in the inertial range.

To derive the most reliable estimate of the large-scale bound of the inertial range, we suggest the following. Ten consecutive wave-numbers along the spectrum were used as a start point to calculate the best linear fit and the slope $\alpha_i$ inside the inertial range. This interval was chosen in such a way that it covers transitioning from the nearly-flat (or slightly inclined) spectrum in the energy injection range to the power-law spectrum in the inertial range (see Figure \ref{fig-2} and 4th column of Table 1. Note that a continuous piecewise linear fitting can be applied for this procedure when a large number of spectra has to be processed). This transition interval is located between two solid vertical line segments marked for each spectrum in Figure \ref{fig-2}. The resulting power index $<\alpha>$ (the last column in Table 1), was derived as an average of $\alpha_i$ (the averaging over ten values of $\alpha_i$ produced the error bars shown in the last column of Table 1), while the middle point of the transition interval was accepted as the left end of the inertial range (the 5th column of Table 1.)

The small-scale bound of the inertial range for HMI data was found to be 2.4-2.2~Mm and was based on  following considerations. The Nyquist frequency of a spectrum is a frequency corresponding to the smallest resolved linear size, $r_{res}$,  which is accepted to be 2 pixels. The MTF of a telescope distorts the high-frequency end of the power spectrum on a rather broad range of scales from $r_{res}$ up to approximately triple resolution limit, as it was discussed in Introduction. We thus accepted the value $r_0 \approx 3r_{res}$ as the small-scale bound of the inertial range. The triple resolution limit of HMI data is $\approx2.2$~Mm. To safely avoid influence of the MTF cutoff at high frequencies, we adopted hereinafter the small-scale bound for HMI spectra as 2.4~Mm.

This definition works well when $r_{res}$ is larger than the diffraction limit of a telescope. Otherwise, we have to accept the diffraction limit as $r_{res}$ to derive $r_0$. In the present study, we limit our spectral analysis to spatial scales larger than $r_0$ and did not apply any MTF correction. 

\section{Results}  

\subsection{Power spectra in undisturbed photosphere from HMI data}

 \begin{figure*}
 	\includegraphics[width=\columnwidth]{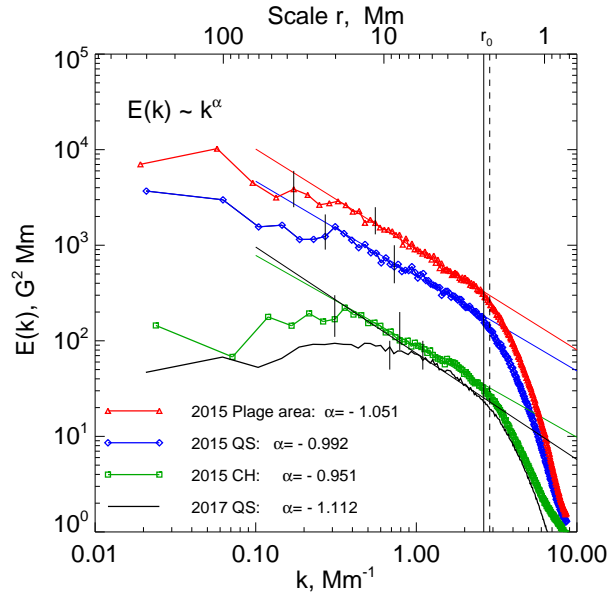}
 	\caption{\sf Magnetic power spectra calculated from the HMI data. 
 		The scale, $r_0$, corresponding to the triple resolution of HMI (2 pixels, or 1\asec) is indicated with the vertical dashed line. The small-scale bound 2.4~Mm of the inertial range is marked by the vertical solid line. For each spectrum, the large-scale transition interval is marked by solid line segments. The average spectral index, $<\alpha>$ and the average linear fit are shown for each spectrum.}
 	\label{fig-2}  
 \end{figure*}
 Magnetic power spectra derived from HMI data as described above are shown in Figure \ref{fig-2}. 
 The spectra show a well pronounced linear (inertial) range on scales above $r_0$ that extends further toward larger scales as magnetic fields in an analysed area increases reaching up to $\sim20$~Mm for magnetic fields of plage areas. The existence of the inertial range indicates the turbulent state and coupling across various spatial scales in magnetized plasma of the photosphere. However, the spectral index, $\alpha$, in $E(k) \sim k^{\alpha}$ is close to -1, which is quite shallower than the index of Kolmogorov turbulence of $-5/3$ \citep[][hereinafter the K41 theory]{Kolmogorov1941}, implying that the observed field distribution can not be entirely attributed to the operation of the direct turbulent cascade (fragmentation of large structures into hierarchy of smaller ones). Instead, additional injection of energy appears to occur inside the inertial range, which is then evenly redistributed along the spectra by means of the inter-scale coupling that includes direct and inverse cascades \citep{Monin1975}. 

In Figure \ref{fig-2} the highest power spectrum (red curve) represents the strongest magnetic fields of the plage region. Note that the averaged over a magnetogram squared magnetic field strength equals to the integral over the spectrum \citep{Monin1975}, so that the spectrum shifts up as the  magnetic environment becomes stronger. Nevertheless all spectra converge at the highest observable frequency, where suppression of power, owing to the MTF and data noise, is nearly the same over the entire undisturbed photosphere. Note that below (7-10)~Mm the shape of the spectra (the curvature and the slope) is identical so that the three curves overlap each other when they are shifted along the y-axis. This implies that the mechanisms responsible for magnetic field transport and generation on scales of 2-10~Mm appears to be the same regardless of the type of the undisturbed photosphere.

\begin{table*}
	\caption{Data sets and parameters of the power spectra}
	\label{tab1}
	\begin{tabular}{lccccc}
		\hline
		Date &  Instrument  & Region& Large-scale bound interval, Mm& Inertial range, Mm & Power index $<\alpha>$  \\  
		
		\hline
		2015 Feb 10 & HMI/SDO& Plage area&36.44 - 12.15& 19.3 - 2.4& -1.051$\pm$0.033\\ 
		2015 Feb 10 & HMI/SDO& QS        & 23.19 - 9.72& 14.4 - 2.4& -0.992$\pm$0.040\\ 
		2015 Feb 10 & HMI/SDO& CH        & 20.23 - 8.48& 12.5 - 2.4& -0.951$\pm$0.057\\
		2017 Jun 19 & HMI/SDO& QS        & 9.20 - 5.95 & 7.4 - 2.4 &-1.112$\pm$0.074\\
		2017 Jun 19 & NIRIS/BBSO& QS     & 4.30 - 1.84 & 3.0 - 0.9 &-1.200$\pm$0.091\\
		\hline
	\end{tabular}
\end{table*}

\subsection{GST/NIRIS quiet Sun magnetic fields}
 
\begin{figure*}
 	\includegraphics[width=\linewidth]{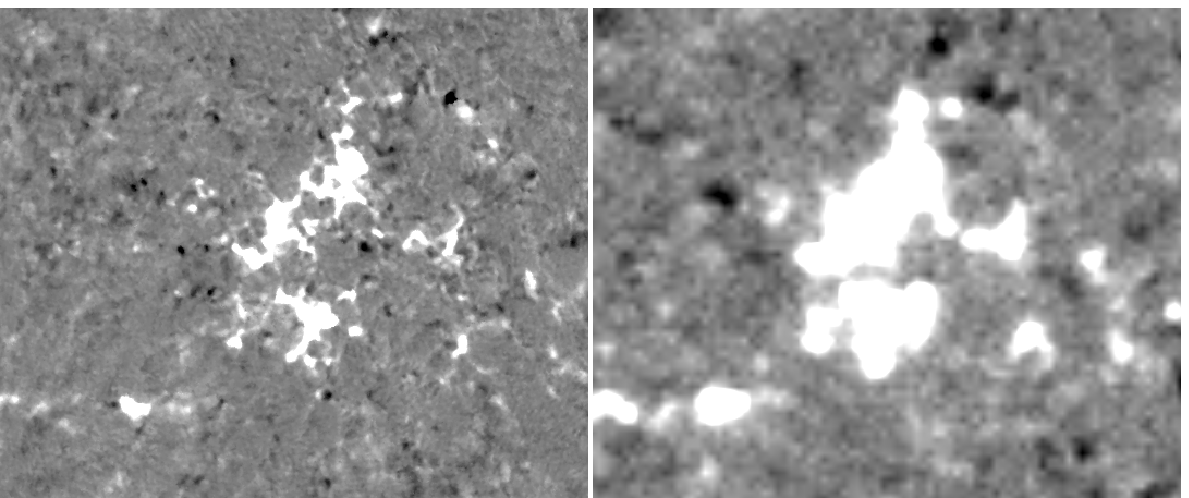}
 	\caption{\sf Left - A GST/NIRIS line-of-sight magnetogram obtained on 19 June, 2017 at 19:19:17~UT at the solar disk center.  The FOV is 540 $\times$ 450 pixels, or 33.3 $\times$ 27.7~Mm.  Right - A co-temporal and co-spatial HMI magnetogram re-scaled to the NIRIS pixel size. The magnetic field strength is scaled between -20~G (black) and 20~G (white) in both magnetograms. }
 	\label{fig-3}  
 \end{figure*}

\begin{figure*}
 	\includegraphics[width=\linewidth]{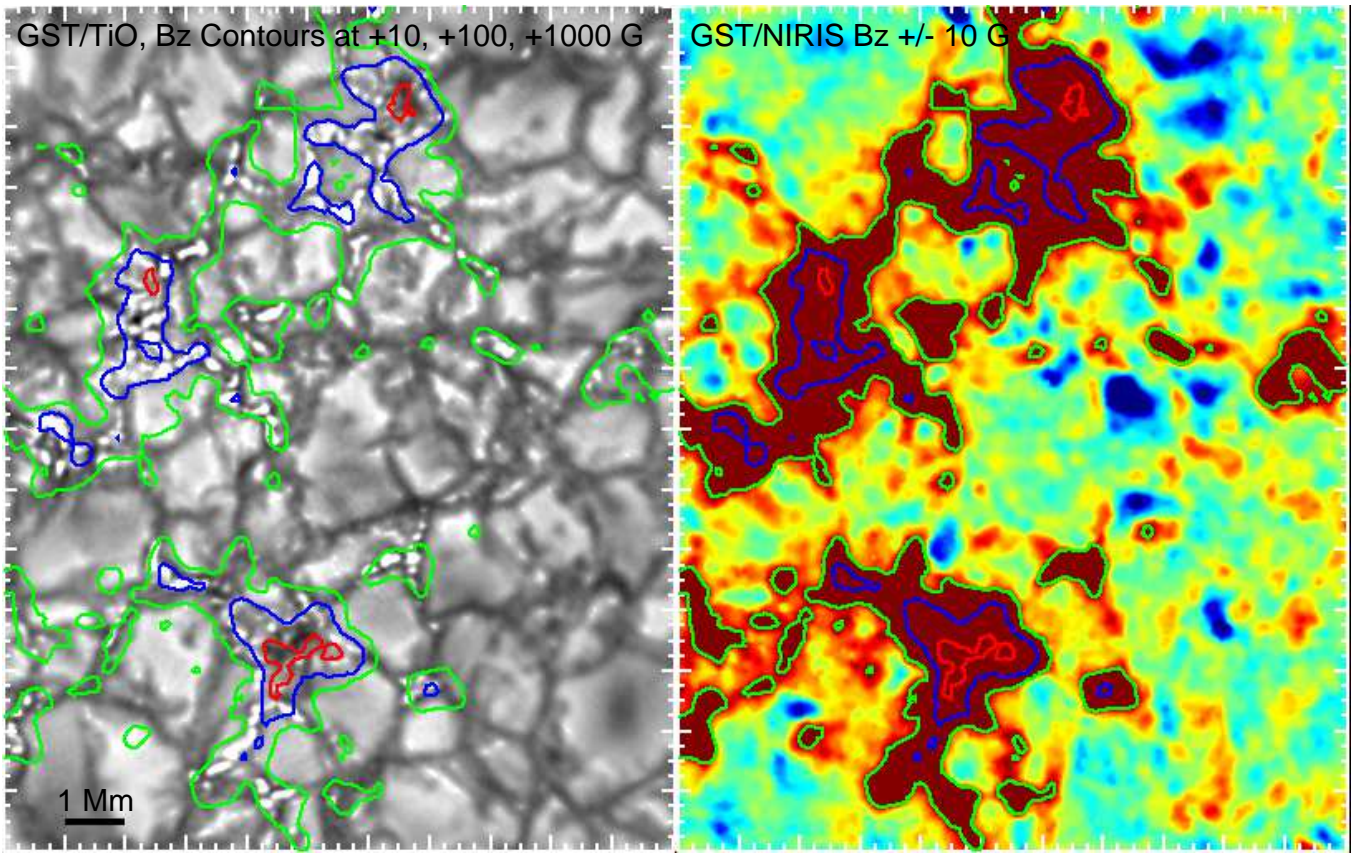}
 	\caption{\sf Left - GST/TiO image overplotted with contours of NIRIS line-of-sight magnetic fields. The green, blue, and red contours correspond to 10, 100, and 1000~G, correspondingly. Right -  NIRIS line-of sight magnetic field (background, the central part of the entire NIRIS FOV shown in Figure \ref{fig-3}, left.) Red/blue represents positive/negative polarities. Large ticks separate 1~Mm intervals. Contours are the same as in the left panel.}  
 	\label{fig-4}  
 \end{figure*}

GST/NIRIS acquired a very good quality QS data set near the disk center on 19 June, 2017. Total 39 sets of full Stokes parameters were recorded between 18:37:46~UT and 19:22:50~UT with a time cadence of 71~s. The best quality magnetogram with the 3$\sigma$ noise level of $\sim5$~G was utilized in this study (Figure \ref{fig-3}, left). In the right panel of Figure \ref{fig-3} we show co-temporal HMI observations of the same area on the Sun. The NIRIS measurements, as compared to the HMI data, show a plethora of unipolar and mixed polarity small-scale magnetic features all over the FOV, where the HMI magnetogram exhibits no or weak, broad, mostly unipolar features. Moreover, the three dominant field concentrations appear more fine structured in the NIRIS data. In Figure \ref{fig-4} we show details of the central part of the NIRIS FOV featuring intense salt and pepper pattern outside the magnetic network as well as several 1~kG clusters (red contours). The co-temporal TiO/GST image (left panel in Figure \ref{fig-4}) demonstrates that these 1~kG fields are associated with clusters of well resolved bright points and dark micro-pores. Strong kilo-Gauss flux tubes are long known to exist across the entire solar surface \citep[e.g.,][]{Stenflo1994}. 
	
\begin{figure*}
 	\includegraphics[width=\linewidth]{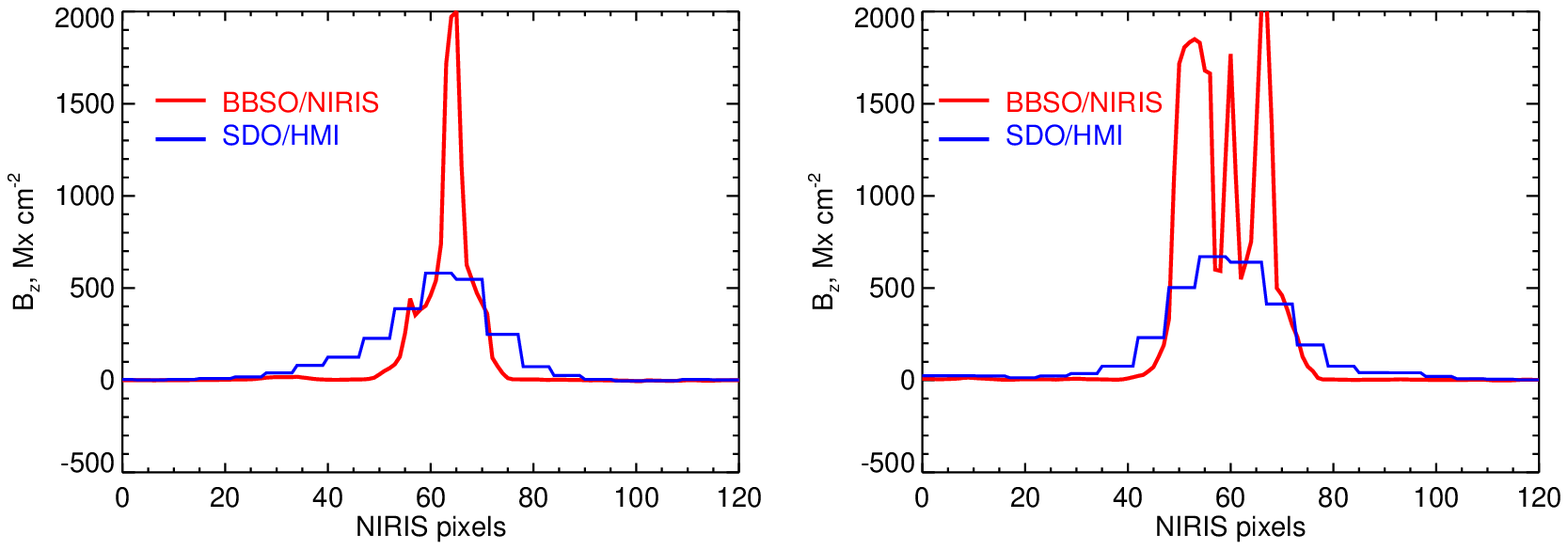}
 	\caption{\sf Variations of  the $B_z$ component along a horizontal cut crossing the upper (left) and lower (right) 1~kG features shown in Fig. \ref{fig-4}. Red/blue curves are NIRIS and HMI profiles.}
 	\label{fig-5}  
 \end{figure*}

In Figure \ref{fig-5} we plot variations of the NIRIS (red) and HMI (blue) $B_z$ component along a horizontal slit cutting across the upper and lower 1~kG elements shown in Figure \ref{fig-4}. As compared to the NIRIS profiles, the HMI-profiles are broader and considerably lower: where the HMI peak is $\approx$ 1500~km wide and registers 600-700~G, the NIRIS resolves one or several $\approx$ 200-400~km wide sub-peaks of 1500-2000~G each. These substructures are wider than the diffraction limit of the NIRIS instrument (170~km). 
	
Our estimation shows that the  NIRIS total unsigned flux over the FOV consists 0.55$\times 10^{20}$ Mx versus 1.0$\times 10^{20}$ Mx for the corresponding HMI data. 
 We note that these are co-temporal measurements of the same area (the same FOV) on the Sun. Some part of this difference is due to differences in instrumentation and calibration techniques. Besides, the noise level of the HMI data is higher than that of the NIRIS data: the 3$\sigma$ noise level is about 18~G for HMI data versus 5~G for NIRIS data. (Note that the total fluxes calculated from pixels above the 3$\sigma$ noise level consist 0.42$\times 10^{20}$ Mx for NIRIS and 0.71$\times 10^{20}$ Mx for HMI data). 
 However,  as we see in Figure \ref{fig-3}, a certain part of this difference is definitely due to better resolution of small magnetic elements in NIRIS data and, therefore, appearance of more extended lacunae of very low magnetic field strength. A presence of few thin kilo-gauss flux tubes is not enough to compensate the overall lowering of the total flux. 
  
\subsection{Power spectra of the quiet Sun from GST/NIRIS data}

\begin{figure*}
 	\includegraphics[width=\columnwidth]{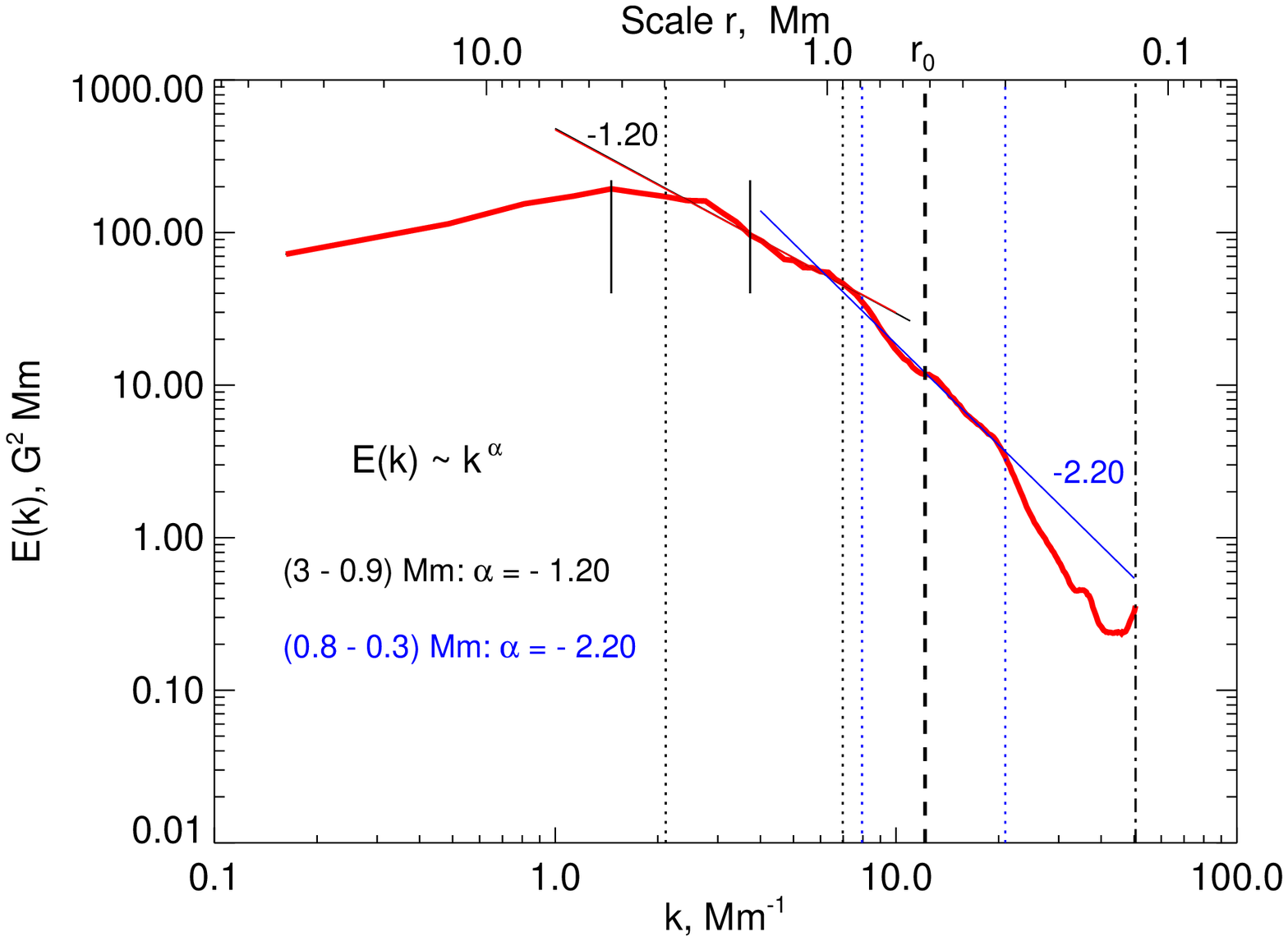}
 	\caption{\sf Magnetic power spectrum derived from the NIRIS magnetogram shown in Figure \ref{fig-3}, left. Vertical dotted lines indicate two linear ranges (shown in parenthesis). The corresponding power indices are listed in the lower left corner of the graph. The vertical solid line segments mark the large-scale transition interval, similar to that presented in Figure \ref{fig-2}. The vertical dashed line at $r_0$ marks the triple resolution limit of GST (which is marked with the dot-dashed line). }
 	\label{fig-6}  
 \end{figure*}

In Figure \ref{fig-6} we plot a magnetic power spectrum derived from the NIRIS magnetogram shown in Figure \ref{fig-3}, left. 

The small-scale bound of the reliable spectral interval, $r_0=510$~km  (dashed line in Figure \ref{fig-6}) in the NIRIS spectrum, was determined by the diffraction limited resolution of the 1.6-meter GST in the spectral range of 15610~\AA, which is 0.17~Mm on the solar surface.

To derive the large-scale bound of the inertial range in NIRIS data we employed the same transition interval approach that we have applied to the HMI data. The interval covers transition from the nearly flat energy injection range to the power-law range and is marked by the vertical solid line segments in Figure \ref{fig-6} and is also listed in the 4th column of Table 1. We then calculated ten estimates of $\alpha_i$ using different starting points and their average $<\alpha>$ is shown in the last column of Table 1. The inertial range, bounded by the middle point of the transition range, is shown in the 5th column. 

Figure \ref{fig-6} shows that the NIRIS spectrum has a small bump on scales of 0.7-1~Mm. Above this range the power index $\alpha$ was found to be -1.20$\pm$0.09, while below this range the spectrum follows the $k^{-2.20\pm 0.05}$ law. Similar power enhancements were also registered in other data such as those shown in Figures 1 and 5 of \citet{Stenflo2012}, where the MTF-correction only slightly elevated this bump but did not eliminate it. \citet{Katsukawa2012} also observed a smooth bump on granular scales in their MTF-corrected spectrum. We did not apply any MTF correction and still we observe this bump. It can not be excluded that the origin of the bump may be related to the existence of a characteristic scale of solar quiet sun magnetoconvection of about 1~Mm, which is a typical scale of granules. Anyway, we did not ignore this feature of the spectrum and defined two linear ranges (3-0.9~Mm and 0.8-0.3~Mm) and calculated their corresponding slopes of -1.2 and  -2.2.
Note that on scales smaller than 1 Mm {\it Hinode} data show a steeper than -2.2  power law. Thus, \citet{Katsukawa2012} reported a slope of -2.7 for the line-of-sight magnetic field spectrum before the MTF-correction, whereas \citet{Stenflo2012} reported a -3.2 slope. 

\subsection{Comparison of NIRIS and HMI spectra}

\begin{figure*}
	\includegraphics[width=\columnwidth]{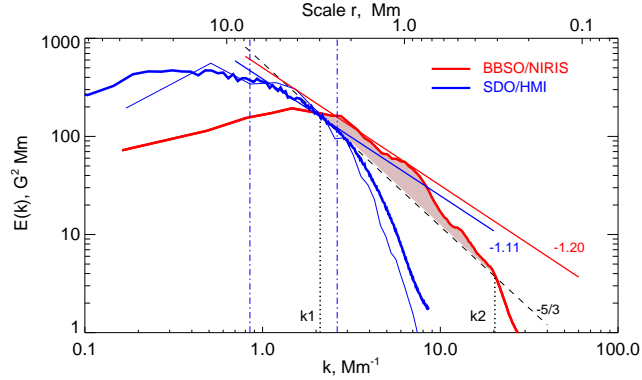}
 	\caption{\sf Comparison of NIRIS and HMI power spectra. Red curve - NIRIS power spectrum and its slope -1.20 defined within the range of 3-0.9~Mm (see 5th row in Table 1). Thin blue curve - the HMI power spectrum calculated over the NIRIS FOV. Thick blue curve - the HMI power spectrum calculated over an extended  FOV of the same magnetogram with the slope -1.11 determined within the range of 7.4-2.4~Mm (marked with vertical dash-dotted line segments, also see the 4th row of Table 1). The black dashed line shows the Kolmogorov-type $-5/3$ spectrum. The shaded area shows the excess of magnetic energy above the $-5/3$ line within the $k_1- k_2$ interval, which contains about 35\% of the NIRIS energy in this interval.}
 	\label{fig-7}  
 \end{figure*}

In Figure \ref{fig-7} we compare NIRIS (red, the same as in Figure \ref{fig-6}) and HMI (thin blue line) power spectra calculated over the same area of an undisturbed photosphere. Because of the limited FOV and rather poor resolution, the HMI spectrum is not smooth and does not 
allow us to make a reliable estimation of the inertial range and the power index. At the same time, the positioning of this spectrum along the $y$-axis is comparable with that of the NIRIS spectrum because they both cover the same area on the Sun. As it was mentioned above (Sec.3.1), the average value of $B_z^2$ determines the height of a spectrum. The relatively strong magnetic features present in the HMI magnetogram with a small FOV (Figure \ref{fig-3}, right) produce a large $<B_z^2>$ thus leading to an elevated spectrum (thin blue line in Fig. \ref{fig-7}). The HMI spectrum derived from an extended magnetogram (930$\times$530 HMI pixels) is positioned along the $y$-axis lower that the NIRIS spectrum mainly because the average $<B_z^2>$ determined over the vast FOV is smaller (see the black line in Fig. \ref{fig-2}). However this spectrum allows us to reliably determine the inertial range and to derive the spectral index. When shifted up along the $y$-axis, this spectrum (thick blue line in Fig. \ref{fig-7}) coincides well with the original HMI-spectrum from the small FOV. Thus spectral parameters for HMI data shown in Fig. \ref{fig-7}, were determined by combining the vertical location of the spectrum derived from the small FOV magnetogram, and the inertial range and the spectral index derived from the large FOV magnetogram.

The plots show that the NIRIS (red) and HMI (blue) linear regression lines are nearly in parallel, and the difference in their slopes (-1.112$\pm$0.074 versus -1.200$\pm$0.091)) is within the error range allowing us to conclude that the spectrum close to $k^{-1}$ may cover a large spatial range starting from nearly 8~Mm and extending toward small scales down to approximately 0.9 Mm.

Figure \ref{fig-7} also demonstrates that on scales smaller than $\approx$3~Mm the magnetic energy registered in NIRIS data is higher than that in HMI data, whereas on larger scales the situation is opposite. We speculate that this is a consequence of the resolution effect as illustrated in Fig.\ref{fig-5}. Indeed, the magnetic elements visible in the HMI magnetograms as monolithic features are seen in NIRIS data resolved into several smaller elements. This fragmentation of magnetic structures is observed on all scales leading to a smooth rearrangement of registered energy toward smaller scales as the resolution improves. This effect explains, at least partially, a depression of the NIRIS spectrum seen on large scales in Fig. \ref{fig-7}. 

If we consider an idea that following the K41 theory energy cascades from large scales (in this case from scales of 3-5~Mm) down to small spatial scales, then the NIRIS spectrum should follow the $-5/3$ line (black dashed line in Figure \ref{fig-7}). However, within the $k_1 - k_2$ interval (3.5 - 0.3~Mm) the observed spectrum is well above the $-5/3$ line, which implies that in addition to the energy cascading down the inertial range, an additional energy may be injected at this spectral range. Our estimates showed that the energy excess constitutes about 35\% of the total NIRIS energy concentrated within the interval of (3.5 - 0.3)~Mm. We emphasise that this is only a lower estimate of the energy excess because future high-resolution observations might reveal a further extension of the shallow spectrum toward the smaller scales.

\section{Conclusions and Discussion}

In summary, an analysis of magnetic power spectra calculated using HMI and GST/NIRIS measurements of magnetic fields in an undisturbed photosphere, allowed us to conclude the following. 

1. All HMI coronal hole, quiet sun, and plage magnetic power spectra have the same spectral index of -1 determined on a broad range of spatial scales covering a nearly decade long interval from 10-20 down to 2.4~Mm. 

2. NIRIS magnetic field data acquired in the magnetic network showed the presence of 200-400 km wide kilo-gauss magnetic elements with the peak field strength of about 2000~G, which are co-spatial with clusters of bright points and dark micro-pores.

3. The HMI magnetic power spectrum $k^{-1.1}$ determined within a range of 7.4-2.4~Mm can be smoothly extended by the NIRIS k$^{-1.2}$ spectrum determined within a range of 3-0.9~Mm. The latter continues with the slope of -2.2 down to 0.3 Mm.
 
4. Comparison of the HMI and NIRIS spectra with the Kolmogorov-type spectrum suggested  that more than 35\% of magnetic energy observed in the scale range of 3.5-0.3~Mm cannot be explained by the Kolmogorov turbulent cascade and should be attributed to other mechanisms of magnetic field generation and/or transport along spatial scales. 

The most plausible explanation for the $k^{-1}$ magnetic spectrum in QS is the local turbulent dynamo acting in the near-surface layers of the convective zone \citep[][]{Vogler2007, Pietarila2010, Rempel2014}. Local dynamo is capable to produce additional (to Kolmogorov turbulent cascade) magnetic energy on scales $\sim$(0.3-10) Mm and thus to make the Kolmogorov-type -5/3 spectrum more shallow and close to $k^{-1}$. Numerical simulations of local turbulent dynamo in the QS photosphere \citep{Danilovic2016} showed that the spectrum of total magnetic energy has an index of -0.73 with a weak tendency for steepening of the spectrum with height.
 The observations presented here show that the $k^{-1}$ magnetic spectrum is observed in different magnetic environments of an undisturbed photosphere. This also agrees with results reported by \citet{Ishikawa2009} that a local dynamo action may generate magnetic fields over the entire solar surface. The power law distribution of magnetic elements have also been interpreted in terms of turbulent dynamo occurring continuously in the convective zone \citep{Parnell2009,Thronton2011}.

The anomalous diffusivity (super-diffusivity) that was found to operate in the photosphere  \citep{Abramenko2011, Lepreti2012, Gianna2013, Gianna2014a, Gianna2014b, Keys2014, Yang2015, Jafar2017, Abramenko2017} may be an additional mechanism that amplifies the action of the local dynamo. Under the super-diffusivity regime, the turbulent diffusion coefficient decreases with decreasing spatial scales thus increasing chances for small magnetic elements to congregate into larger magnetic structures, some of which may be further intensified via convective collapse to become photospheric bright points with kilo-gauss fields. Although the kilo-gauss fields alone can not account for the observed excess of energy observed in the NIRIS power spectrum. By slowing down the rate of energy cascade from large to small scales, super-diffusivity may facilitate accumulation of additional energy at the small-scale end of the spectrum thus making the $-5/3$ spectrum shallower.

\section*{Acknowledgements}

SDO is a mission for NASA Living With a Star (LWS) program. The SDO/HMI data were provided by the Joint Science Operation Center (JSOC). BBSO operation is supported by NJIT and US NSF AGS-1821294 grant. GST operation is partly supported by the Korea Astronomy and Space Science Institute (KASI), Seoul National University, the Key Laboratory of Solar Activities of Chinese Academy of Sciences (CAS), and the Operation, Maintenance and Upgrading Fund of CAS for Astronomical Telescopes and Facility Instruments.
Calculations and analysis of power spectra (V.A.) were supported by Russian Science Foundation grant 18-12-00131. V.Y. acknowledges support from NSF AST-1614457, AFOSR FA9550-19-1-0040 and NASA 80NSSC17K0016, 80NSSC19K0257, and 80NSSC20K0025 grants.
Authors are thankful to Referee for useful comments and suggestions helping much to improve the paper.

\section*{Data availability}

The HMI data that support the findings of this study are available in
the  Joint Science Operations Center (JSOC)
(http://jsoc.stanford.edu/) and can be accessed under open for all
data policy.

Raw NIRIS data were generated at the Big Bear Solar Observatory of
NJIT and are available from one of the co-authors (V.Yu.) on request.

Derived NIRIS data products supporting the findings of this study are
available from the corresponding author V.A. on request.




\bibliographystyle{mnras}



\bsp	
\label{lastpage}
\end{document}